# Langmuir probe diagnostic studies of pulsed hydrogen plasmas in planar microwave reactors

A. Rousseau[*,1], E. Teboul[2], N. Lang[3], M. Hannemann[3], J. Röpcke[3]

[1] Laboratoire de Physique des Gaz et des Plasmas, CNRS, Université Paris-Sud - Bat. 210, 91405 Orsay Cedex, France.
[2] JY/Horiba, Inc., 3880 Park Avenue, Edison, NJ 08820, USA
[3] Institut für Niedertemperatur-Plasmaphysik, Friedrich-Ludwig-Jahn-Str. 19, 17489 Greifswald, Germany.
* Corresponding author: Antoine.Rousseau@lpgp.u-psud.fr

**ABSTRACT**

Langmuir probe techniques have been used to study time and spatially resolved electron densities and electron temperatures in pulse-modulated hydrogen discharges in two different planar microwave reactors ($f_{microwave}$= 2.45 GHz, $t_{pulse}$= 1 ms). The reactors are (i) a standing-wave radiative slotted waveguide reactor and (ii) a modified travelling-wave radiative slotted waveguide reactor, which generate relatively large plasmas over areas from about 350 cm$^2$ to 500 cm$^2$. The plasma properties of these reactor types are of particular interest as they have been used for basic research and for plasma processing, e.g. for surface treatment and layer deposition. In the present study the pressures and microwave powers in the reactors were varied between 33 and 55 Pa and 600 and 3600 W, respectively. In regions with high electromagnetic fields shielded Langmuir probes were used to avoid disturbances of the probe characteristic. Close to the microwave windows of the reactors both the electron density and the electron temperature showed strong inhomogeneities. In the standing-wave reactor the inhomogeneity was found to be spatially modulated by the position of the slots. The maximum value of the electron temperature was about 10 eV and the electron density varied between 0.2 and 14×10$^{11}$ cm$^{-3}$. The steady state electron temperature in a discharge pulse was reached within a few tens of microseconds whereas the electron density needed some hundreds of microseconds to reach a steady state. Depending on the reactor the electron density reached a maximum between 80 and 200 µs after the beginning of the pulse.

# 1. INTRODUCTION

Non-equilibrium plasmas are used in a variety of applications, e.g. for thin film deposition, semiconductor processing, surface activation and cleaning. These plasmas are of growing interest not only in the field of plasma processing, because of their favourable properties, but also in basic research. Discharges excited by microwaves are characterized by an efficient power coupling without the use of electrodes. This design leads to advantages, in particular when high purity discharges are needed. Typical for microwave plasmas containing molecular gases such as hydrogen, is a relatively high degree of dissociation which in turn can lead to high chemical reactivity.

In the last decade several different approaches have been made to create microwave plasmas over extended volumes needed for plasma processing. In the low pressure range (0.1-1 Pa) magnetic confinement has lead to highly ionised plasmas in relatively large volumes with or without electron cyclotron resonance heating [1-3]. At higher pressures (10-100 Pa) the decrease of the mean free path length of the electrons causes an inhomogeneous density of the plasma for large volumes or surfaces [4]. One general approach to overcome the problem of inhomogeneity consists of one or two waveguides combined with slots or antennas in such a way that the electromagnetic field coupled into the reactor volume is consistently homogenous at a certain distance from the waveguide. Such large sized reactors have been constructed in circular [8-9] and planar geometry [2, 5-9].

Over the last few years planar microwave plasma reactors have been used extensively in a continuous mode for plasma chemical applications [5], for example, in diamond deposition, surface corrosion protection by deposition of organosilicon compounds, surface cleaning, and plasma induced surface modification [6,10-12]. The analysis of excitation processes such as the determination of the gas temperature and of the degree of dissociation in plasmas containing molecular feed gases, e.g. hydrogen or diborane, was also examined using emission spectroscopy [13-15]. Recently, this type of reactor has been used to investigate the chemistry and kinetics in plasmas with admixtures of hydrocarbons and other reactive gases by infrared absorption spectroscopy [16,17].

An separate field using microwave plasmas in non-equilibrium conditions is that using non-stationary power incoupling, for example, by pulse excitation or power modulation. In principle, non-stationary excitation conditions caused by time dependent energy supply into the plasma can be used to manipulate the density and energy distribution of electrons and to achieve modified chemical equilibria [18,19]. As a result with pulsed plasmas higher chemical conversion rates but also improved etching and deposition processes can be realized [20-24]. The thermal influence on sensitive substrates can be controlled by changing the duty cycle.

The investigation of plasma physics and chemistry *in situ* requires detailed knowledge of excitation and relaxation phenomena and of plasma parameters, which can be obtained by appropriate diagnostic techniques complemented by modelling approaches.

Recently the time evolution of rotational, vibrational and kinetic temperatures using emission spectroscopy from pulsed $H_2$ microwave plasmas generated in a planar travelling-wave reactor was reported [25]. In microwave surface wave discharges the degree of dissociation of $H_2$ was found to be much higher under pulsed conditions (1kW peak power) than in a continuous mode (1 kW cw) for the same power [26]. Hence modelling of a $H_2$-$CH_4$ microwave discharge showed that the concentration of different hydrocarbon radicals can be controlled by duty cycle ratio [27]. Ashida *et al.* studied the breakdown in pulsed argon discharges. The authors described how the high electric field at breakdown leads to an enhancement of the ionisation frequency, and therefore to an increase of the electron density compared to continuous mode operation [28]. Spatial analysis of the plasma ignition in a circular radiative slotted waveguide reactor showed that, even when the electron density is nearly uniform at the steady state, the discharge ignites in very specific places [29]. In contrast to the many studies of pulsed RF plasmas, systematic time resolved investigations of pulse-modulated microwave discharges, in particular concerning properties of the electron component, are relatively uncommon. The need for a better scientific understanding of phenomena in pulsed molecular microwave plasmas must be based on an improved knowledge of plasma ignition including the typical time behaviour of plasma density formation.

The present article describes temporally and spatially resolved Langmiur probe diagnostic studies of electron density and electron temperature in pulsed modulated hydrogen discharges in two different *planar* microwave reactors ($f_{microwave}$= 2.45 GHz), (i) in a standing-wave radiative slotted waveguide reactor and (ii) in a modified travelling-wave radiative slotted waveguide reactor [5].

Although the Langmuir probe technique is an intrusive method it provides one of the few approaches for acquires information about the electron parameters in the plasma. In particular, in numerous low pressure DC and RF discharges electrical probes have been successfully applied. Several robust theories have been proposed since the early work of Langmuir in the 1920's [30]. The bases of modern theories were set by the work of Allen *et al*. [31] and Bohm [32] and extended by Laframboise [33] and Chen [34]. None of these theories take into account ion-neutral collisions across the sheath. However, under the experimental conditions relevant in this study, the electrostatic sheath surrounding the Langmuir probe is collisionally dominated. For this reason the analysis of the experimental data in this paper is referred as to the collisional probe theory, proposed by Zakrzewski *et al*. [35].

## 2. EXPERIMENTAL

### 2.1. Plasma reactors

The experiments were performed in two different planar reactors; a standing-wave slotted waveguide reactor (abbreviated as the SWG reactor), Figure 1(a), and a modified travelling-wave slotted waveguide reactor (abbreviated as the TWG reactor), Figure 1(b), [5]. In both cases, the microwave generators used worked in pulsed mode and were externally triggered by a pulse generator. The length of a whole period was chosen as 10 ms for the experiments with the SWG reactor and 50 ms for the TWG reactor. Under both experimental conditions the pulse length was 1 ms.

The general properties of linear radiating applicators, guiding the microwave energy into the plasma chamber, have been described in detail elsewhere [5,7,36]; only a brief account is given here.

*SWG reactor:*

The applicator of the SWG reactor is shown in Figure 1(a). It consists of a waveguide (length 54 cm) terminated by a shorting circuit that establishes a standing wave. The microwave power is injected into the plasma chamber via 5 resonant slots that act as dipole radiators. Each slot is 6.1 cm long (half the microwave wavelength in vacuum) and is separated from another slot by half a microwave wavelength in the waveguide (8.7 cm). The angle between the slot axis and the waveguide axis is 20°. The radiating waveguide is located 3 cm above a quartz window of 9.3 cm width that closes the plasma chamber. The plasma is generated directly under this quartz window. The grounded walls of the plasma chamber are made of copper and stainless steel. The reactor width is 13 cm. A microwave generator, Sairem GMP 12KE, provided 600 W of power in the pulse with a rise time of 40 μs. The incident power was zero between the pulses. The surface of the plasma region in the plane below the quartz window is typically about 350 –400 cm$^2$. Details of the pulse regime can be seen in Figure 2.

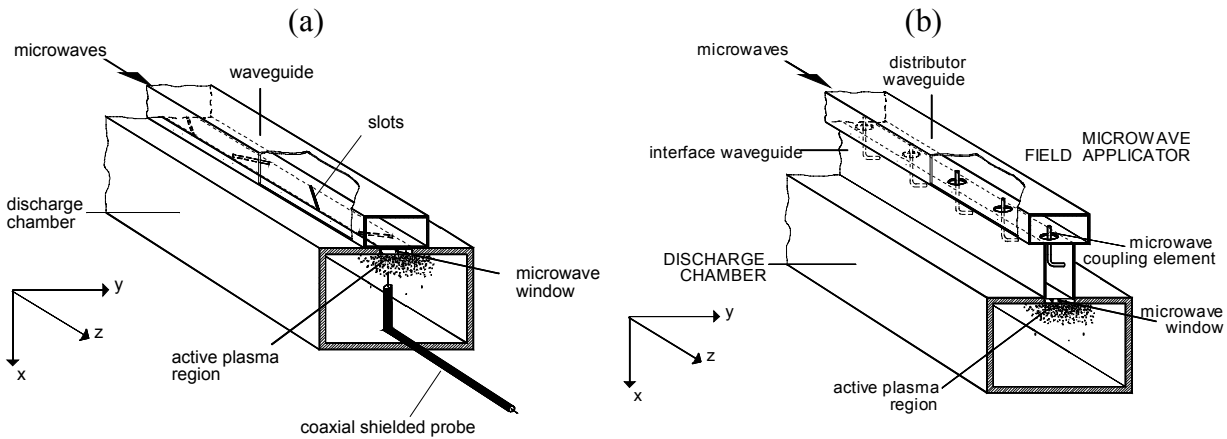

**Figure 1:** The different planar microwave reactors: The applicator of the SWG reactor consists of 5 slots (1a), the applicator of the TWG reactor of 26 holes with coupling elements and an additional smoothing interface waveguide (1b).

*TWG reactor:*
Because the applicator of the TWG reactor takes advantage of having a local distributed and tuneable coupling of the microwave power combined with an additional smoothing, considerably homogeneous plasma excitation can be achieved over a large surface, typically up to 500 cm$^2$. A scheme of the T-shape configuration of the applicator consisting of two rectangular waveguides, the distributor waveguide and the interface waveguide, is shown in Figure 1(b). The discharge chamber has dimensions of 25*20*140 cm$^3$ (W*H*L). The microwave power is radiated via 26 coupling holes separated from each other by a distance of $\lambda/4$ (=3.8 cm). In order to tune the power and to improve the coupling efficiency these holes contain moveable metallic hooks. At each end of the interface waveguide a matched load leads to the existence of travelling waves. The dimensions of the waveguides are those of type R22 (a:b=2:1 with a=11cm). A quartz window, which allows the microwave power to penetrate into the discharge chamber, forms the fourth mechanical wall of the interface waveguide. When a plasma is generated it represents the fourth conducting wall of the interface waveguide.

A microwave generator, Sairem 60KE/DC, provided 3.6 kW of power in the pulse for a rise time of between 20 and 40 µs. The minimum power between the pulses was 600 W.

**2.2. Langmuir probe system**
A single, cylindrical Langmuir probe was used for electron density and temperature determinations. The probe was mechanically moveable in the discharge reactors (see Figure 1) and electrically controlled by an acquisition system, DIGIPROBE obtained from the Jobin-Yvon Horiba Group. The scan range of the probe voltage was ± 190 V and the maximum probe current was limited to 100 mA. The DIGIPROBE acquisition system permits time resolved electron density measurements to be made with a minimum time resolution of 30 µs. A TTL pulse, synchronised with the microwave pulse, triggers the system. After the trigger pulse the probe tip is polarised with an adjustable delay related to the microwave pulse, and the probe current measured (Figure 2). The probe voltage is then reset to 0 V in order to avoid ohmic heating of the probe, which can occur in higher density (10$^{12}$ cm$^{-3}$) plasmas. The current measurement at the same probe voltage can be repeated over many pulses in order to average the signal. The probe characteristic is achieved by varying the polarisation (typically from –50 to +25 V).

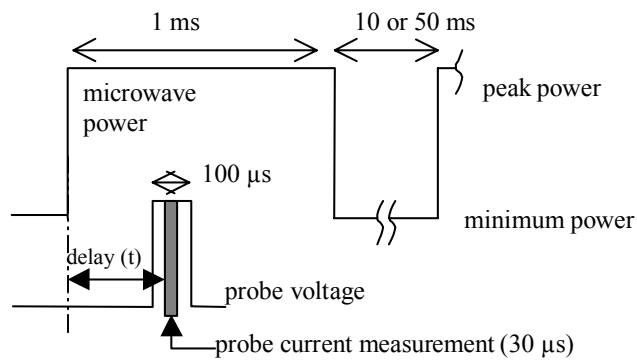

**Figure 2:** Schematic diagram of the microwave pulse and of the time resolved probe acquisition procedure.

For pulsed measurements, it is often necessary to average many probe characteristics in order to obtain a good signal to noise ratio (typically 20-100). Ordinary cylindrical Langmuir probes made of a tungsten tip inside a dielectric capillary gave no relevant results in the microwave reactors used in this study, since these probes are strongly perturbed by the electromagnetic field of the microwaves. Therefore, a special shielded probe made of a micro coaxial arrangement was used. It consisted of a copper layer on floating potential surrounding its dielectric of 1 mm radius. The tip radius and length are 0.25 mm and 3-7 mm, respectively. Figure 3 presents characteristics of two different probes. One was recorded using an ordinary tungsten tip inserted into an aluminium capillary, and the other was obtained by the shielded probe. In the latter the electron current increases exponentially with the applied voltage, up to the plasma potential, and then saturates. The unshielded probe characteristic shows completely different behaviour. The electronic part is no longer exponential but linear and increases without any saturation like the classical characteristics of a DC discharge.

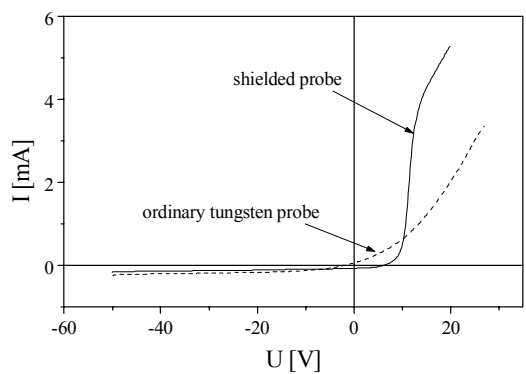

**Figure 3:** Probe characteristics recorded in the SWG reactor at 33 Pa and 600 W using a shielded probe (solid line) and an ordinary tungsten/alumina probe (dashed line).

The use of the shielded micro coaxial probe has lead to an enormous improvement because electromagnetic perturbation can be completely eliminated in the case of the SWG reactor. However, some perturbation of the probe characteristic still remained in the TWG reactor, caused by the essentially higher peak power of 3.6 kW.

In the TWR reactor the Langmuir probe could be moved in 3 dimensions (x, y, z). In the SWG reactor, only axial (z axis) and vertical measurements (x axis) have been performed. Scans along the x axis were carried out at two distances from the quartz window: x = 0.3 cm and x = 2.5 cm.

In both reactors the hydrogen flow rate was measured on mass flow controllers. The total flow rate was kept constant at a few sscm to avoid introducing gas impurities into the discharge chamber. The pumping speed was adjusted with a butterfly valve to maintain a constant pressure in the plasma and the pressure measured on a capacitance pressure gauge. The pressures were 33 Pa and 55 Pa for the SWG and TWG reactors respectively.

## 3. PROBE THEORIES AND ANALYSIS PROCEDURE

Since the present experimental study was mainly focused on the mapping of two different large volume plasma reactors, and concerned with the internal distributions of the time dependence of the plasma density and the mean energy of the electrons, an analysis of electron energy distribution functions was not performed. Further more it is not intended to discuss various electrostatic probe theories in this paper, and only a brief account of the fundamental principles is given here.

The plasma density is deduced from the ion saturation current, mainly because the ionic branch of the electrostatic probe characteristic is less sensitive to electromagnetic disturbances than the electron one. In the present work the pressures are relatively high, 33 and 55 Pa. Additionally, the relatively large probe radius of 0.25 mm leads to collisions of ions with neutrals, since the width of the sheath is directly related to the probe radius. Therefore a suitable collision theory is used as described below.

Several theories have dealt with the problem of analysing the probe current since the early work of Langmuir [30]. Two major theories are often used:

(i) Assuming conservation of both angular momentum and ion energy Laframboise developed a self-consistent theory of ion collection by cylindrical and spherical probes. This theory used the two-body central-force equations in order to determine orbits of the ions for different values of $T_i / T_e$ ($T_i$: ion temperature, $T_e$: electron temperature) [33]. Within this theory some of the ions that enter the probe sheath are only deflected by the attractive potential and escape without reaching the probe. Laframboise theory holds as long as no ion collisions inside the sheath occur, which is the case for very low pressures and high plasma densities.

(ii) The other theory is based on the assumption that cold ions ($T_i / T_e = 0$) move radially toward the probe, which means that all ions entering the probe sheath reach the probe tip. It was shown by Allen *et al.* that this is valid for spherical probes but not for cylindrical ones [31]. Nevertheless, Chen used this assumption also for cylindrical probes [34]. Measurements with such probe geometry may often be interpreted with more accuracy by Chen's theory than by the approach of Laframboise, due to destruction of the orbital motion of the ions by collisions. Actually, this "radial motion" theory is valid for a cylindrical probe as long as the number of collisions across the sheath is small (typically less than one), but not zero.

As noted above, in our experiments the gas pressure was 33 and 55 Pa depending on the reactor. Under these circumstances the sheath has to be considered as no longer free of collisions. As shown in the following section, typical values of the plasma density are in the range of $2 \times 10^{10}$ to $6 \times 10^{11}$ cm$^{-3}$ and of the electron temperature from 2 to 7 eV. Assuming a sheath width of ten times the Debye length, the sheath width ranges from 300 to 500 μm. Thus, collisions and momentum transfers of $H_3^+$ with $H_2$ are likely, since the mean free path length of $H_3^+$ is about 130 μm at 55 Pa and 1000 K, which is smaller than the width of the electrostatic sheath. Therefore, the theory excluding collisions should be corrected in order to take this effect into account. A parametric synthesis of these two theories has been proposed by Zakrzewski *et* *al*. [35] for cylindrical probes collecting positive ions. The ion current I of this theory is derived from the Laframboise ion current $I_L$ by

$$I = \gamma_1 \gamma_2 I_L \qquad (eq.\ 1)$$

Where $\gamma_1$ is the rate of increase of the Laframboise ion current due to destruction of the orbital motion of the ions and is evaluated from Chen's ion current. The parameter $\gamma_2$ is the rate of decrease of the Laframboise ion current by scattering of ions out of the probe sheath according to Schultz and Brown [37] and Jakubovski [38]. For details of the ion density determination according to this theory see [39,40].

In the present study the electron temperature is deduced from the slope of the second derivative of the electron current (total probe current minus extrapolated ionic current). In all experiments reported in this paper the following relationship

$$K_e(K_e+1)(K_e+\ln(l_p/r_p))^{-1} > 1 \qquad (eq.\ 2)$$

was valid ($K_e = \lambda_e/r_p$ - electron Knudsen number with $\lambda_e$ electron mean free path, $r_p$ probe radius, $l_p$ probe length):
As a consequence, it was possible to extract the plasma potential from the voltage value, where the second derivative of the total probe characteristic is zero, even when the sheath surrounding the probe is not free of collisions [41]. The accuracy of the ion density or electron temperature determined by the methods described above is no better than 20 % [39].

## 4. RESULTS AND DISCUSSION

### 4.1. Spatial plasma density gradients at the stationary state
#### *4.1.1. SWG reactor*

We first discuss the spatial profiles of $N_e$ and $T_e$ in the stationary state, that is, at the end of the 1 ms pulse (Figure 2). Figure 4 shows axial distributions of (a) the electron density and (b) of the electron temperature, at two different distances from the quartz window in the SWG reactor. The delay between the beginning of the pulse and the measurements was set to 850 µs in order to achieve a steady state (pulse length: 1ms, pulse period: 10 ms)

The axial electron density profile is not homogeneous; two strong maxima are evident at z = 23 cm and z = 33 cm. A third seems to be located very close to the power source (z ≈ 10 cm). Each maximum occurs between two slots. It is clear that the electron density profiles are very similar at 0.3 and 2.5 cm from the quartz window. This means that axial diffusion along the z axis does not lead to a smoother profile of the electron density. Close to the quartz window (the area of the penetrating microwaves) the minimum electron density (z = 18 and 27 cm) is about $2\times10^{10}$ cm$^{-3}$ and the maximum density reaches $1.5\times10^{11}$ cm$^{-3}$ (7 times higher). This can be correlated with the electron temperature profile (Figure 4(b)). As expected, the electron density and temperature profiles have their maxima at the same positions. The electron temperature varies very strongly from 2 eV to 7 eV. However, in contrast to the electron density distribution, the electron temperature is very sensitive to the distance from the quartz window. The lower position (2.5 cm) shows that $T_e$ is much less modulated (only between 2 and 3.5 eV). This means that the microwave electric field, which is very strong and non homogeneous close to the quartz window is absorbed by the plasma. The electric field intensity decreases rapidly with increasing distance from the quartz window x.

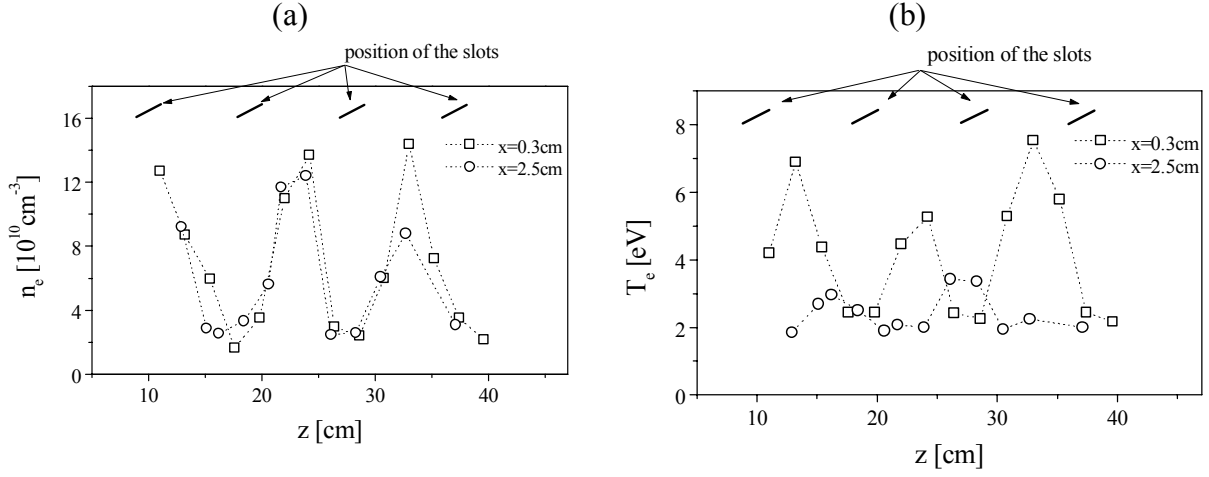

**Figure 4:** SWG reactor: steady state axial electron density profile (4(a)) and electron temperature profile (4(b)) at two different distances from the quartz window (33 Pa, 600 W, t = 850 μs).

The electron density gradient is most probably due to the electron energy gradient. Actually, a rigorous description of the discharge particle and energy balance would require 3D modelling, which is beyond the scope of this paper. However, it is of interest to roughly estimate the spatial variation of the ionisation frequency due to the electron energy gradient. Assuming a Maxwellian electron distribution function the ionisation frequency can therefore be expressed as,

$$\nu_i = \frac{n_{H_2}}{(\pi m_e)^{1/2}} \left(\frac{2}{kT_e}\right)^{3/2} \int_{\varepsilon_i}^{\infty} \sigma_i(\varepsilon)\varepsilon \exp\left(\frac{-\varepsilon}{kT_e}\right) d\varepsilon \qquad \text{(eq. 3)}$$

where $\varepsilon_i$ is the ionisation threshold of $H_2$ (16.5 eV), and $m$, $\varepsilon$ and $T_e$ are the electron mass, energy and temperature respectively. The ionisation cross section may be approximated by: [42]

$$\sigma_i(\varepsilon) = \sigma_0 \cdot \frac{\varepsilon - \varepsilon_i}{\varepsilon_i} \qquad \varepsilon > \varepsilon_i$$
$$\sigma_i(\varepsilon) = 0 \qquad \varepsilon \leq \varepsilon_i \qquad \text{(eq. 4)}$$

The corresponding ionisation frequency is:

$$\nu_i = n_{H_2} \sigma_0 \overline{v} \left(1 + \frac{2kT_e}{\varepsilon_i}\right) e^{\left(-\frac{\varepsilon_i}{kT_e}\right)} \qquad \text{(eq. 5)}$$

Where v is the mean velocity $\overline{v} = (8kT_e/\pi m_e)^{1/2}$. Close to the quartz window $T_e$ ranges from 2 to 7 eV so that the corresponding ionisation frequency should be multiplied by a factor of 1000. As mentioned above, the plasma density is only multiplied by a factor 7 due to diffusion.

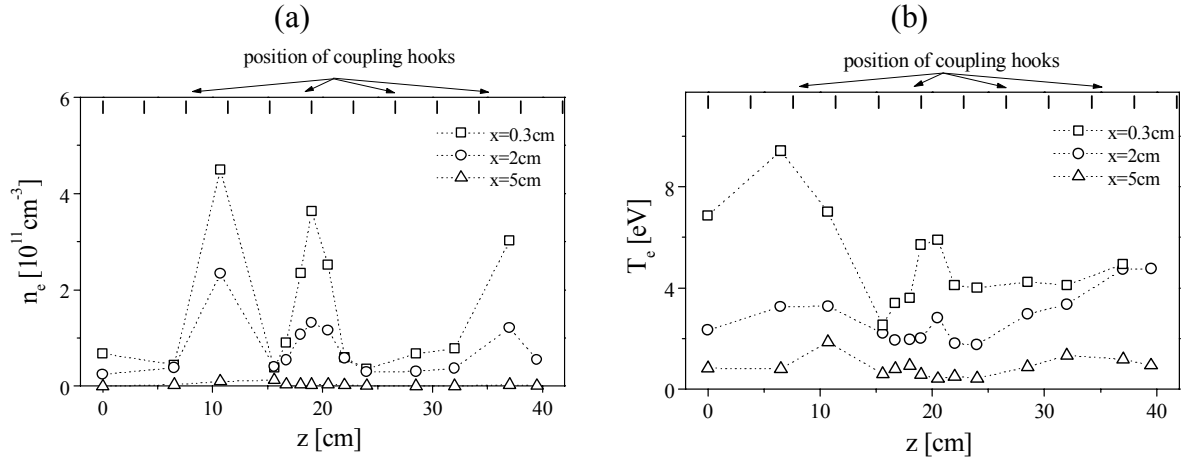

**Figure 5:** TWG reactor: steady state axial electron density profile (5(a)) and electron temperature profile (5(b)) at three different distances from the quartz window (55 Pa, 3600 W, t = 850 μs).

### 4.1.2. TWG reactor
*Axial measurements:*

Considering now the results obtained in the TWG reactor. Figure 5(a) and 5(b) show the **axial** plasma density and temperature profiles respectively at the steady state, at three different distances from the quartz window. Operating conditions are: 55 Pa, 3.6 kW, t = 850 μs. The plasma density profiles are also as strongly modulated as in the case of the SWG reactor. The plasma density ranges from $5 \times 10^{10}$ to $4.5 \times 10^{11}$ cm$^{-3}$, which is higher than the values measured in the SWG reactor. However, the plasma density decreases as the distance from the quartz window increases, in contrast to the situation in the SWG reactor. This is most likely due to the higher pressure, 55 Pa instead of 33 Pa, leading to reduced electron diffusion. At a distance of 2 cm from the quartz window the electron density shows a decay by a factor of about 2, and at 5 cm, it is reduced by a factor of more than 20. This means that the plasma itself (i.e. the collection of charged particles) only spreads over about 3 cm.

Similarly to the SWG reactor case, the electron energy decreases markedly with distance from the quartz window (Figure 5(b)). A very high electron energy, near to 10 eV, is measured at z = 6 cm. This can be related to the strong electric field generated by the high peak power (3.6 kW). However, it is important to note that in such a high field region the electron part of the probe characteristic is disturbed.

*Transverse measurements:*

Figure 6 shows profiles along the *y* axis of the TWG reactor. The distance from the quartz window was 0.3 cm. Two different positions along the z axis are considered: a bright plasma zone and a dark plasma zone (z = 20.5 cm and 15.2 cm respectively according to Figure 5). Spatial distributions of the plasma density and electron energy are presented in Figures 6(a) and 6(b) respectively. Measurements were performed ± 3 cm from the reactor axis. In the scanned region the electron density is approximately constant whereas $T_e$ is constant only in the dark zone and exhibits a strong radial profile in the bright zone.

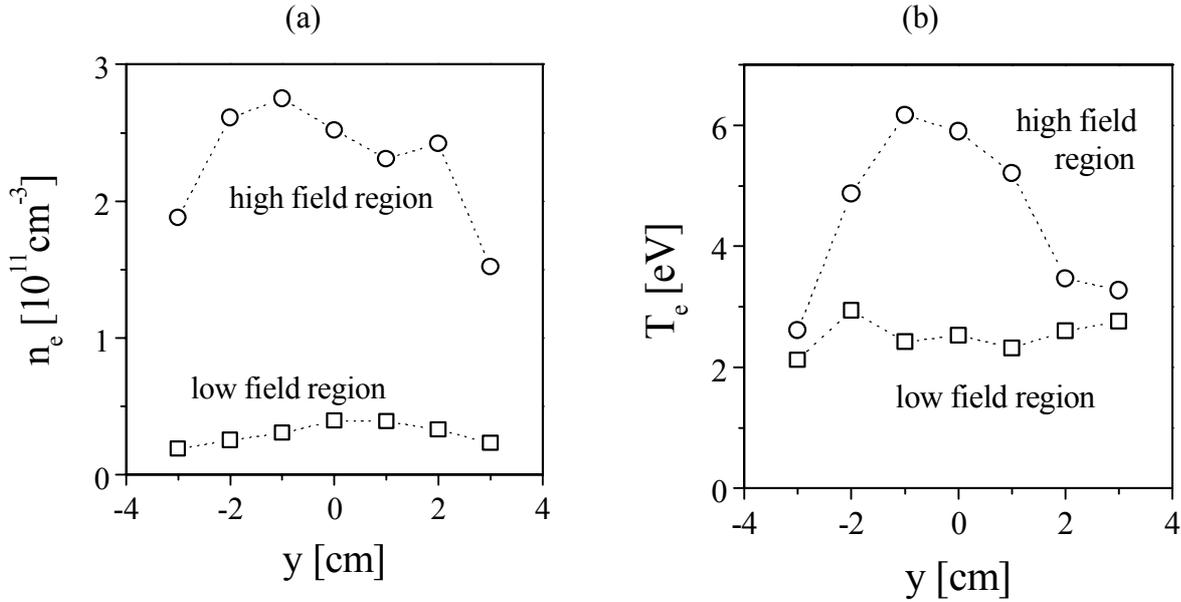

**Figure 6:** TWG reactor: steady state transverse electron density profile (6(a)) and electron temperature profile (6(b)) at two different axial positions z = 20.5 cm (high field region) and z = 15.2 cm (low field region). The position y = 0 corresponds to the centre of the quartz window (55 Pa, 3600 W, x = 0.3 cm, t = 850 μs).

### 4.2. Time resolved profiles measurements
#### *4.2.1. Time evolution of $n_e$ and $T_e$ axial profiles in the SWG reactor*

Figure 7 presents the time evolution of the axial electron density profiles in the SWG reactor close to the quartz window (x = 0.3 cm) at four different pulse times (t = 0, 50, 150 and 850 μs). This clearly shows the ignition behaviour of the plasma. At t = 0 μs, the plasma density is close to zero everywhere except close to the microwave power injection region (z < 10 cm). Some 50 μs later the plasma density is close to its maximum value over 75 % of the reactor volume but still remains low for z > 30 cm.

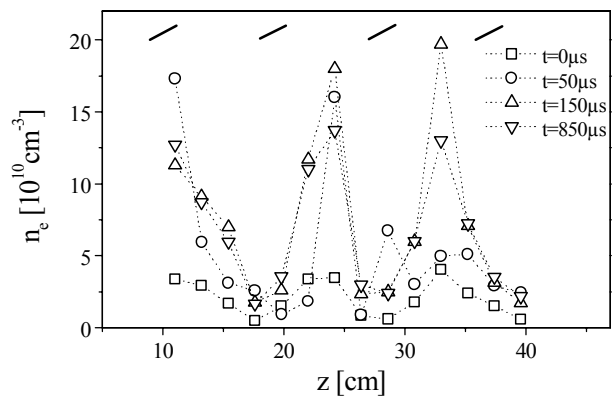

**Figure 7:** Time evolution of the electron density profile in the SWG reactor (33 Pa, 600W, x = 0.3 cm).

These effects illustrate the propagation of the ionisation front. At t = 150 μs, the plasma density has reached its maximum value in the whole volume. Figure 7 also shows the electron density profile at its steady state (t = 850 μs), as presented on Figure 4(a). It is interesting to note that in high field regions (z = 10, 23 and 33 cm) the electron density is higher at t = 150 μs than at t = 850 μs.

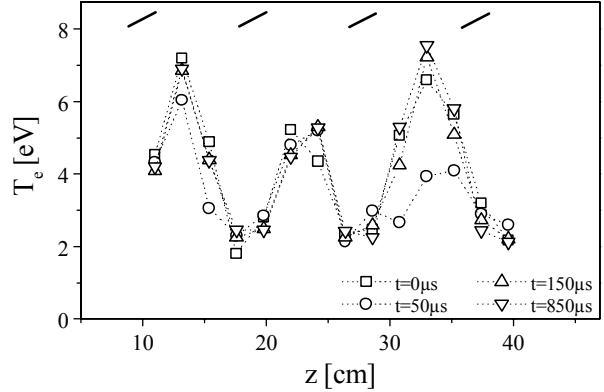

**Figure 8:** Time evolution of the electron temperature profile in the SWG reactor (33 Pa, 600W, x = 0.3 cm).

Figure 8 shows the temporal evolution of the electron temperature under the same conditions as Figure 7. We clearly see, that, even at the beginning of the pulse, the electron temperature has almost reached its steady state (except at the end of the reactor (z > 30 cm)).

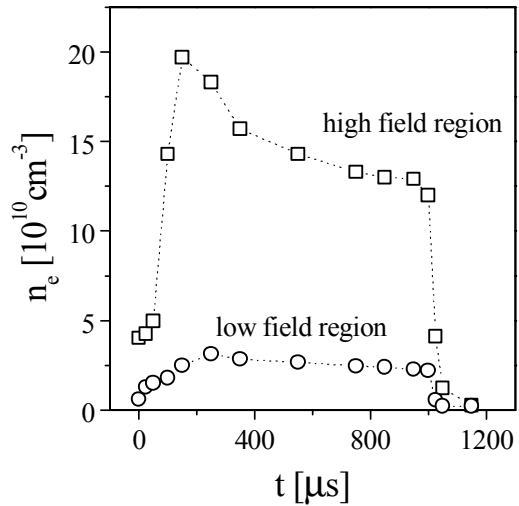

**Figure 9:** Time evolution of the electron density for two different positions z = 33 cm (high field region) and z = 17.5 cm (low field region) in the SWG reactor (33 Pa, 600W, x = 0.3 cm).

In order to discuss the time evolution of both electron density and temperature the time evolution of the electron density at two axial positions z = 33 cm (high field region) and z = 17.5 cm (low field region) is shown in Figure 9. The experimental conditions were the same as in Figure 7 and 8. In Figure 10 the time evolution of the electron temperature is given. As mentioned above, $T_e$ reaches its stationary state almost immediately, which is too short to be observed with the probe system. The overvoltage causes the maximum of the electron temperature during the breakdown. In fact, it is well known that when an electromagnetic field is applied to a neutral gas, the breakdown electric field is much higher than that required to maintain the plasma in the steady state. Consequently, a high electron temperature during a short period of time is observed.

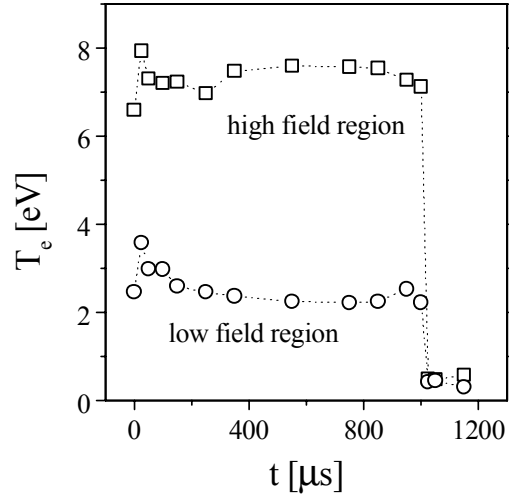

**Figure 10:** Time evolution of the electron temperature for two different positions z = 33 cm (high field region) and z = 17.5 cm (low field region) in the SWG reactor (33 Pa, 600W, x = 0.3 cm).

On the other hand, the temporal evolution of the plasma density is much slower: $N_e$ reaches a maximum value in about 200 μs and then decreases smoothly to reach a steady state around 500μs. The maximum of the electron density can be explained in two different ways. (i) 200μs after the beginning of the pulse, the volume of the plasma has not yet reached its steady state. Then, the power density is higher at t = 200 μs than during the stationary state after the expansion of the plasma, which might explain the maximum in the plasma density. (ii) Taking a thermal effect into consideration, the injected microwave power density is high enough to substantially heat the gas.

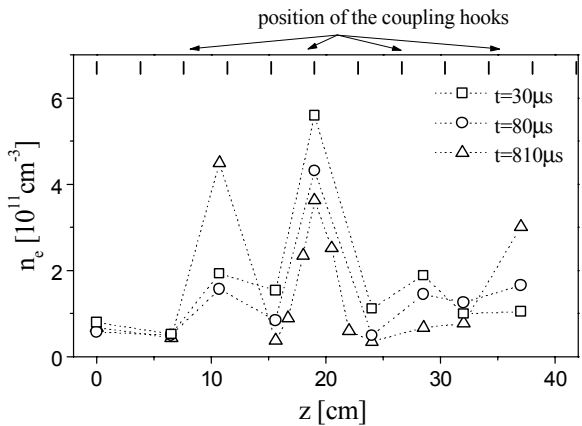 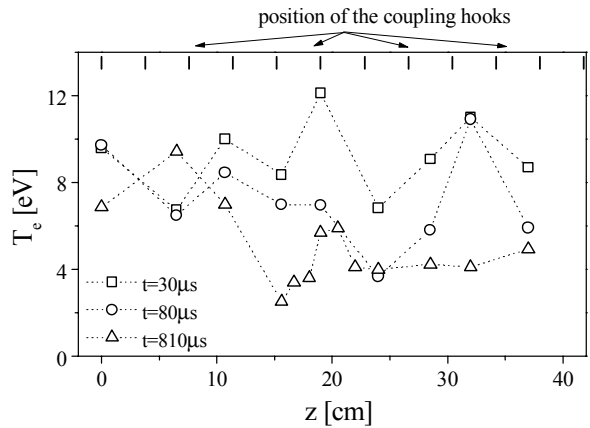

**Figure 11:** Time evolution of the electron density axial profile in the TWG reactor at different time points after the front edge of the pulse (x = 0.3 cm). At t = 810 μs the stationary state is established.

**Figure 12:** Time evolution of the electron temperature axial profile in the TWG reactor at different time points after the front edge of the pulse (x = 0.3 cm). At t = 810 μs the stationary state is established.

Experiments performed in the TWG reactor showed that the gas temperature increased by approximately 600 K within 500 µs when the power was 3.6 kW [25]. Such gas heating leads to a decrease of the neutral particle density as well as of the ionisation frequency, which might also explain the maximum of the electron density at t = 200 µs.

### 4.2.2. Time evolution of $n_e$ and $T_e$ axial profiles in the TWG reactor

Similar time evolutions of electron density and temperature profiles were also found in the TWG reactor. Figure 11 shows the time evolution of the axial plasma density profile (z axis) at t = 30, 80 and 810 µs. At t = 810 µs the stationary state is reached. The time evolution of the plasma density in the high and low field region is shown in Figure 13. In the high field region, the maximum plasma density value is reached within 60 µs. This is shorter than that measured in the SWG reactor. This is mainly due to using a minimum power of 600 W between two pulses to prevent the plasma density falling to zero. Hence when the 3.6 k W peak power is applied there already exists a weak plasma which allows the ionisation front to propagate faster. It is also interesting to note that in contrast to that found in the SWG reactor (Figure 9) the plasma density does not fall to a zero value in the TWG reactor. In the SWG reactor, the plasma density almost reaches zero for a decay time of 30µs, which is comparable to the decay time of the microwave generator itself. The presence of a maximum of the plasma density at t = 1200 µs (200 µs after the end of the pulse) is discussed in the following section.

### 4.2.3. Time evolution of $n_e$ and $T_e$ post discharge

The residual plasma between two pulses, caused by the remaining 600 W of microwave power, is apparent in Figure 14. The time evolution of the electron temperature in this figure shows a decrease from 6 to 2 eV, which is still a relatively high value. In the case of the SWG reactor, the power is zero between two pulses, and $T_e$ drops to 0.5 eV in less than 30 µs (Figure 10).

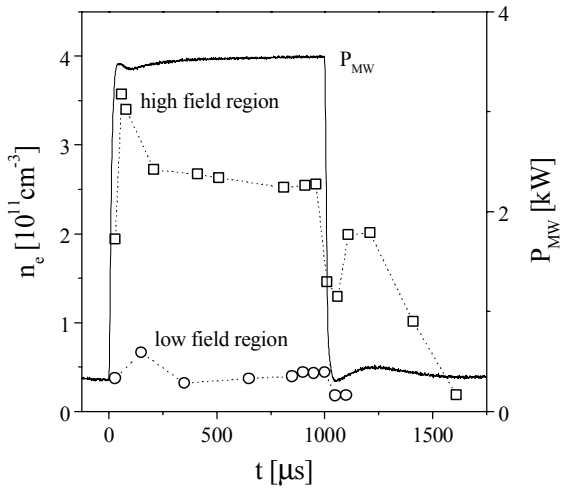 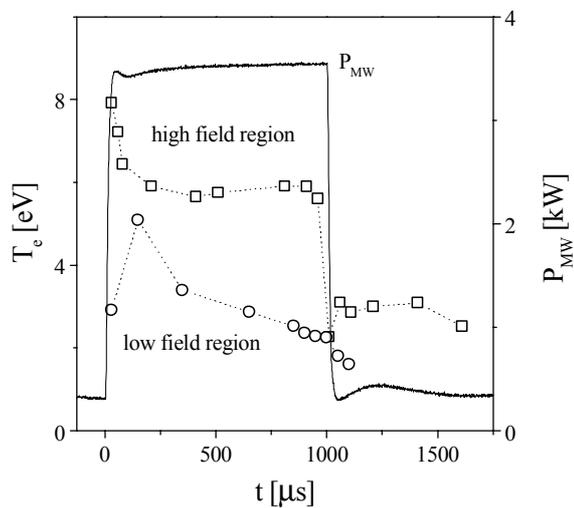

**Figure 13:** Time evolution of the electron density for two different positions z = 20.5 cm (high field region) and z = 15.2 cm (low field region) in the TWG reactor (55 Pa, 3600W, x = 0.3 cm).

**Figure 14:** Time evolution of the electron temperature for two different positions z = 20.5 cm (high field region) and z = 15.2 cm (low field region) in the TWG reactor (55 Pa, 3600W, x = 0.3 cm).

Concerning the maximum of the electron density, 200 μs after the end of the pulse (Figure 13), we suppose that this is due to the coupling (feed back) between the plasma density and the injected power: When the microwave power is decreased from 3600 to 600 W, the decay of the plasma density leads to a time dependence for the plasma impedance and the injected power.

It is interesting to note that the amplitude of the electron density is more pronounced when the collisional theory of Zakrzewski and Kopinczynski [35] rather than the classical non-collisional theory from Chen [34] is used. This interesting fact is actually due to the time evolution of the gas temperature, and the related neutral density during the pulse and post discharge. Under slightly different experimental conditions, (same peak power, same gas pressure, but 50 % instead of 10 % duty cycle ratio), the gas temperature was measured both from Doppler broadening of $H_2$ Fulcher lines, and from population density distributions of molecular levels [25,43]. It was shown that the gas temperature is about 600 K between pulses (due to the residual 600 W input power), whereas it increases to 1050 K during the 3.6 kW pulse itself. Moreover, post discharge, the gas temperature drops with a time constant of about 30 μs. Therefore, the neutral particle density increases during the first 50 μs, which increases the number of collisions and decreases the correction factor $\gamma_2$ in eq. 1. Furthermore the factor $\gamma_1$ gives an increase of the Laframboise current in eq. 1 only in the region $0 < r_p / \lambda_D \leq 3$ where $\lambda_D$ is the Debye length. The ratio $r_p / \lambda_D$ has a value of about 7 at this specific time in the afterglow. This explains why Chen's non-collisional probe theory underestimates the maximum electron density mentioned above.

## 5. CONCLUSIONS

Time and spatially resolved plasma density and electron temperature measurements have been performed in two different linear plasma reactors. These experiments were made possible by the use of a shielded Langmuir probe, in order to reduce electromagnetic disturbances. For discharge pressures near 50 Pa, a probe theory which includes ion/neutral collisions across the sheath has to be applied.

It was found that in both reactors the plasma density is strongly inhomogeneous close to the microwave window. Variations from 5 to 10 have been recorded between high density and low density regions. Away from the microwave windows, the plasma density decreases and becomes more homogeneous due to charged particle diffusion.

Time resolved density measurements show that the electron density reaches a steady state within about 500 μs. It goes through a maximum 100 to 200 μs after the beginning of the pulse. Compared to the SWG reactor the steady state is reached earlier in the TWG reactor, probably because of the residual plasma which remains between two pulses in this reactor. The time evolution of the spatial plasma density profile shows that the plasma ignites preferentially at places near to the microwave power injection points and then propagates from these.

Although this is the first time that electron densities and temperatures have been analysed with both temporal and spatial resolution in two planar microwave reactors, it is clear that the extension of the present studies to the measurement of electron energy distribution functions is desirable for a more comprehensive understanding of the excitation phenomena in these plasmas. This includes the verification of the measured parameters of the electrons by other independent methods, e.g. microwave interferometry or Thompson scattering, to improve the modelling potential of pulsed molecular microwave plasmas.

**Acknowledgements:** We are indebted to the Deutscher Akademischer Austauschdienst and EGIDE for support of this project as part of the French-German PROCOPE Collaboration

Program (Project 9822831). In addition, the project was partly supported by the Deutsche Forschungsgemeinschaft, Sonderforschungsbereich 198. Dr. P.B. Davies provided helpful discussions and S. Saß and D. Gött valuable technical assistance for which we are grateful.

## REFERENCES

[1] M. Pichot, A. Durandet, J. Pelletier, Y. Arnal and L. Vallier, Rev. Sci. Instrum. **59** (1988), p. 1072
[2] N. Sato, S. Iizuka, Y. Nakagawa and T. Tsukada, Appl. Phys. Let. **62** (1993), p. 1469
[3] T. Lagarde, J. Pelletier and Y. Arnal, Plasma Sources Sci. Technol. **6** (1997), p. 56
[4] M. Moisan and J. Pelletiers, Microwave excited plasmas, Elsevier, Amsterdam, 1992.
[5] A. Ohl, Microwave Discharges : Fundamentals and Applications, Ed. C. M. Ferreira and M. Moisan, NATO ASI series, Series B: Physics 302 (1992), p. 205
[6] A. Ohl, H. Strobel, J. Röpcke, H. Kamerstetter, A. Pries and M. Schneider, Surface Coatings Technol. **74-75** (1995), p. 59
[7] A. Ohl, Journal de Physique IV **8** (1997), p. Pr7-83.
[8] F. Werner, D. Korzec and J. Engemann, Plasma Sources Sci. Technol. **3** (1994), p. 473
[9] D. Korzec, F. Werner, R. Winter and J. Engemann, Plasma Sources Sci. Technol. **5** (1996), p. 216
[10] A. Ohl and J. Röpcke, J Diam. Relat. Mater. **1** (1992), p. 243
[11] A. Ohl, J. Röpcke and W. Schleinitz, Diam. Relat. Mater **2** (1993), p. 298
[12] J. Röpcke, A. Ohl and M. Schmidt, J. Analyt. Spectrometry **8** (1993), p. 803
[13] A. S. Astashkevich, M. Käning, E. Käning, N. V. Kokina, B. P. Lavrov, A. Ohl and J. Röpcke, J. Quant. Spectrosc. Radiat. Transfer **56** (1996), p. 725
[14] B. P. Lavrov, A. S. Melnikov, M. Käning and J. Röpcke, Phys. Rev. E **59** (1999), p. 3562
[15] M. Osiac, B. P.Lavrov and J. Röpcke, J. Quant. Spectrosc. Radiat. Transfer. (2002) in press.
[16] J. Röpcke, L. Mechold, M. Käning, W. Y. Fan and P. B. Davies, Plasma Chem. Plasma Process. **19** (1999), p. 395
[17] F. Hempel, L. Mechold and J. Röpcke, XV. ESCAMPIG, Lillafüred, Miskolc 2000, Z. Donko, L. Jenik, J. Szigeti Eds., Europhys. Conf. Abstr. **24F** 64
[18] S. Sumakawa and S. Furuoya, Appl. Phys. Lett. **63** (1993), p. 2044
[19] H. Sugai, K. Nakamura, Y. Hikosaka and M. Nakamura, J. Vac. Sci. Technol. A**13** (1995), p.887
[20] H. Chatei, J. Bougdira, M. Remy, P. Alnot, C. Bruch and J. K. Krüger, Diam. Relat. Mater. **6** (1997), p.505
[21] A. Hatta, H. Suzuki, K. Kadota, H. Makita, T. Ito and A. Hiraki Plasma Sources Sci. Technol. **5** (1996), p.235
[22] R. Boswell and K. Henry, Appl. Phys. Lett. **47** (1985), p.1095
[23] S. Samukawa and T. Tsukada, J. Vac. Sci. Technol. A **15** (1997), p.391
[24] J. Behnisch, F. Mehdorn, A. Holländer and H. Zimmermann, Surf. Coat. Technol. **98** (1998), p.875
[25] N. Lang, M. Kalatchev, M. Käning, B. P. Lavrov and J. Röpcke, Proc. Frontiers in Low Temperature Plasma Diagnostics III, p. 253, Saillon, Switzerland (1999)
[26] A Rousseau, L Tomasini, G Gousset, C Boisse-Laporte and P Leprince J. Phys. D: Appl. Phys. **27** (1994) p. 2439
[27] K Hassouni, X. Duten, A. Rousseau, A. Gicquel and M. H. Gordon, Plasma Sources Sci. Technol. **10** (2001), p. 61
[28] S. Ashida, C. Lee, M. A. Lieberman, J. Vac. Sci. Technol. **Vol. A** (1995), p. 2498
[29] A. Brockhaus, S. Behle, A. Georg and J. Engemann, Journal de Physique IV **8** (1997), p.

[30] I. Langmuir, H. M. Mott-Smith, Gen. Elect. Rev. **26** (1923), p. 731
[31] J. E. Allen, R. L. F. Boyd, P. Reynolds, Proc. Phys. Fluids **2** (1957), p. 112.
[32] D. Bohm, "The characteristic of electrical discharges in magnetic field", ch.3, Ed. Guthrie and Mc Graw Hill, Book Co. Inc., 1949
[33] J. G. Laframboise, Univ Toronto. UTIAS Rept. N° 100, 1966.
[34] F. F. Chen, " Electric Probes ", in Plasma Diagnostic Techniques, Ed. R. H. Huddlestone and S. L. Leonard, Academic Press, New York, 1965
[35] Z. Zakrzewski, T. Kopiczynski, Plasma Phys. **16** (1974), p. 1194.
[36] Z. Zakrzewski, M. Moisan, Plasma Sources Sci. Technol. **4** (1995), p. 379.
[37] G. J. Schulz, S. C. Brown, Phys. Rev. **98** (1955), p. 1642.
[38] A. K. Jakubowsky, AIAA Journal, **8** (1972), p. 988
[39] P. David, M. icha, M. Tichy and T. Kopiczynski, Contr. Plasma Phys. **30** (1990), p. 167
[40] O. Chudacek, P. Kundra, J. Glosik, M. Sicha and M. Tichy, Contr. Plasma Phys. **35** (1995), p. 503
[41] S. Klagge, M. Tichy, Czech. J. Phys., **B35**, (1985), p. 988
[42] M. A. Liebermann and A. J. Lichtenberg, Principles of plasma discharges and materials processing, Wiley & Sons, New York, 1994.
[43] N. Lang, B. P. Lavrov and J. Röpcke, Plasma Sources Sci. Technol. (2002), to be submitted.